# The Effect of Magnetic Turbulence Energy Spectra and Pickup Ions on the Heating of the Solar Wind

C. S. Ng[a], A. Bhattacharjee[b], P. A. Isenberg[b], D. Munsi[b], and C. W. Smith[b]

[a]*Geophysical Institute, University of Alaska Fairbanks, Fairbanks, AK 99775, USA.*
[b]*Space Science Center, University of New Hampshire, Durham, NH 03824, USA*

**Abstract.** In recent years, a phenomenological solar wind heating model based on a turbulent energy cascade prescribed by the Kolmogorov theory has produced reasonably good agreement with observations on proton temperatures out to distances around 70 AU, provided the effect of turbulence generation due to pickup ions is included in the model. In a recent study [Ng et al., J. Geophys. Res., **115**, A02101 (2010)], we have incorporated in the heating model the energy cascade rate based on Iroshnikov-Kraichnan (IK) scaling. We showed that the IK cascade rate can also produce good agreement with observations, with or without the inclusion of pickup ions. This effect was confirmed both by integrating the model using average boundary conditions at 1 AU, and by applying a method [Smith et al., Astrophys. J., **638**, 508 (2006)] that uses directly observed values as boundary conditions. The effects due to pickup ions is found to be less important for the IK spectrum, which is shallower than the Kolmogorov spectrum. In this paper, we will present calculations of the pickup ions effect in more details, and discuss the physical reason why a shallower spectrum generates less waves and turbulence.

## INTRODUCTION

Since the proton temperature in the solar wind is observed to decrease with heliocentric distance slower than predicted by adiabatic expansion, it is believed that an *in situ* source is required to heat the solar wind [1,2]. Good agreement with the observed temperature profile has been obtained using a quasi-steady solar wind turbulence evolution model [3-6] that includes turbulence generation due to pickup ions in the outer heliosphere [6-10]. Later developments of this model include extensions to the case with nonzero cross-helicity [11-13].

The heating of the solar wind in the model is provided by the dissipation of turbulent energy, which cascades from large to small scales, and is eventually dissipated at the dissipation scale. Since in steady state, the heating rate is essentially the same as the energy cascade rate in the inertial range, the precise functional form of the cascade rate is an important ingredient of the model. Although different forms of the cascade rate were considered in the early development of the model [3,14], based on the Kolmogorov theory of hydrodynamic turbulence as well as the Iroshnikov–

Kraichnan (IK) theory of incompressible magnetohydrodynamic (MHD) turbulence, later work involving detailed comparisons with observations was done assuming the Kolmogorov cascade rate. In a recent paper, we have investigated the consequences of Kolmogorov or IK scaling on the problem of proton heating, assuming that the turbulence is isotropic [15], and have shown that the IK scaling can also give predictions consistent with observations. In the model using Kolmogorov scaling, the contribution from pickup ions is important for larger distance from the sun. However, the heating from the model becomes too much at the outer heliosphere. In [15] we have shown that the pickup ion contribution using IK scaling is considerably smaller and thus seems to produce better agreement at the outer heliosphere. We will concentrate on discussing this effect due to pickup ions in this paper.

## PICKUP IONS EFFECTS IN SOLAR WIND HEATING MODEL

The solar wind heating model discussed above is derived based on several strong and simplifying assumptions (see [13] and other references therein). Among them are a steady and a spherically symmetric solar wind, an isotropic Kolmogorov scaling, a constant radial solar wind speed $V_{SW}$, and a constant Alfvén speed $V_A (<< V_{SW})$. Under these conditions, the evolution of solar wind turbulence as a function of the heliocentric distance $r$ can be modeled by the following set of equations [8-10,16], with terms in the right hand side to the left of the arrow:

$$\frac{dZ^2}{dr} = -\frac{A'}{r}Z^2 - \frac{\alpha Z^3}{\lambda V_{SW}} + \frac{Q}{V_{SW}} \rightarrow -\frac{A'}{r}Z^2 - \frac{\alpha Z^4}{\lambda V_{SW} V_A} + \frac{Q}{V_{SW}}, \quad (1)$$

$$\frac{d\lambda}{dr} = -\frac{C'}{r}\lambda + \frac{\beta Z}{V_{SW}} - \frac{\beta \lambda Q}{V_{SW} Z^2} \rightarrow -\frac{C'}{r}\lambda + \frac{\beta}{V_{SW}}\left(\frac{Z^4}{V_A}\right)^{1/3} - \frac{\beta \lambda Q}{V_{SW} Z^2} \quad (2)$$

$$\frac{dT}{dr} = -\frac{4T}{3r} + \frac{m\alpha Z^3}{3k_B \lambda V_{SW}} \rightarrow -\frac{4T}{3r} + \frac{m\alpha Z^4}{3k_B \lambda V_{SW} V_A}. \quad (3)$$

In this model, the turbulence is characterized by two quantities: the average fluctuation energy (in Elsässer units) $Z^2 = \langle \delta v^2 + \delta b^2/4\pi\rho \rangle$, where $\rho = nm$ is the solar wind density ($n$ and $m$ are proton density and mass respectively), and the correlation length of the fluctuations, $\lambda$. Note that by describing the turbulence energy by only one field, $Z^2$, we have assumed zero cross-helicity, i.e., $Z_+^2 = Z_-^2 = Z^2$, where $Z_\pm^2 = \langle \mathbf{Z}_\pm^2 \rangle$ with $\mathbf{Z}_\pm = \delta \mathbf{v} \pm \delta \mathbf{b}/(4\pi\rho)^{1/2}$. The constant parameters $A'$ (negative) and $C'$ are used to model the effect of stream compressions and shear. The function $Q$ (with the functional form given below) represents the fluctuation source due to interstellar pickup protons. The constant parameters $\alpha$ and $\beta$ are estimated from considerations of local turbulence theory [4,14]. The factor $Z^3/\lambda$ in the second term on the right hand side of Eq. (1) or (3) is due to the Kolmogorov cascade rate (see discussion below in this section). Here $T$ is the solar wind proton temperature, which evolves passively according to Eq. (3) but does not affect the evolution of $Z^2$ and $\lambda$. Note that the first term on the right hand side of Eq. (3) describes the adiabatic cooling due to the expansion of the solar wind, while the second term represents the heating due to dissipation of the turbulent energy.

In the Kolmogorov theory, the energy $\delta v^2$ of the scale $\lambda$ is estimated to cascade to the next scale in an eddy turnover time $\tau \sim \lambda/\delta v$. Therefore, the energy cascade rate is $\varepsilon \sim \delta v^2/\tau \sim \delta v^3/\lambda$. On the other hand, in the IK theory of MHD turbulence, the energy cascade becomes $\varepsilon \sim \delta v^4/\lambda V_A$. To examine the effect of using IK cascade, we have simulated the model, using terms in the right hand side of Eqs (1) to (3) to the right of the arrow. Note that the factor $Z^4/\lambda V_A$ is due to the IK energy cascade rate. Although the pickup ions terms involving $Q$ appear formally unchanged, they too depend on the assumed energy cascade rate, as described below.

The functional form of $Q$ in this model is $Q = \zeta(V_{SW}^2/n)(dN/dt)$, where $\zeta$ is the fraction of newly ionized pickup proton energy that generates waves. Here $V_{SW}^2/n$ is the initial kinetic energy per pickup proton in the same units as $Z^2$ in the plasma frame, and $dN/dt$ is the rate at which pickup protons are created, which can be modeled by the equation $dN/dt = N_0 v_0 (r_E/r)^2 \exp(-L/r)$, where $L$ is the scale of the ionization cavity, $N_0$ is the neutral hydrogen density at the termination shock, and $v_0$ is the ionization rate at $r = r_E = 1$ AU.

Following [9] and [16], the factor $\zeta$ is calculated from the equation

$$\zeta(\Delta) = 1 - \frac{\Delta + V_{SW}^{-4}\int_{\Delta}^{1} v^4(\mu)S(\mu)d\mu}{\Delta + V_{SW}^{-2}\int_{\Delta}^{1} v^2(\mu)S(\mu)d\mu}, \tag{4}$$

where $\Delta = Z/3^{1/2}V_A$, $v(\mu)$ is the solution obtained by integrating the equation,

$$\frac{dv}{d\mu} = \left[\sum_j \frac{V_j I_j(k_r)}{|\mu v - W_j|}\left(1 - \frac{\mu V_j}{v}\right)\right]\left[\sum_j \frac{I_j(k_r)}{|\mu v - W_j|}\left(1 - \frac{\mu V_j}{v}\right)^2\right]^{-1}, \tag{5}$$

subject to the initial condition of $v(\mu = \Delta) = V_{SW}$, $S(\mu)$ is a scale factor calculated by taking the difference between $v(\mu)$ and another solution of Eq. (5) using the initial condition $v(\mu = \Delta) = 1.001 V_{SW}$, normalized to $S(\mu = \Delta) = 1$. Here $V_j$ and $W_j$ in Eq. (5) are the phase and group velocity of the $j$th wave mode resonating with the cyclotron resonant wave number $k_r = \Omega/(\mu v - V_j)$, where $\Omega$ is the proton cyclotron frequency. Using the cold plasma dispersion relation $\omega(k) = \pm k V_A (1 + \omega/\Omega)^{1/2}$, $V_j$ can be obtained by solving the third-order equation $V_j^3 - \mu v V_j^2 + \mu v V_A^2 = 0$, and $W_j$ is given by $W_j = -2\mu v V_A^2/(2V_j^2 - 2\mu v V_j + V_A^2)$.

Note that there is only one resonant wave mode if $|\mu v| < 1.5\sqrt{3}V_A$, and three modes otherwise. The function $I(k)$ in Eq. (5) is determined by the one-dimensional energy spectrum of the turbulence. When the energy spectrum is Kolmogorov, $I(k)$ is given by $I(k) = A(r)|k|^{-5/3}$. The function $A(r)$ does not enter the final result since it is cancelled in Eq. (5) at each position $r$. For the study using the IK scaling, we need to use the IK spectrum instead, i.e., $I(k) = A(r)|k|^{-3/2}$.

The coefficients $A'$ and $C'$ in these two sets of equations can in principle be different, depending on the spectral index, and this variation may change the model predictions significantly. However, since we estimate them by dimensional arguments (e.g., see [13]), which do not depend on the spectral index explicitly, we will choose values for $A'$ and $C'$ that are the same as those used in previous studies [8-10,16], in order to have a meaningful comparison with earlier results.

To see how different spectral indices affect the value of $\zeta$, which depends only on dimensionless parameters $\Delta$ and $V_A/V_{SW}$ according to Eqs. (4) and (5), we have plotted color-coded contours of its values as a function of these two parameters in Figure 1 (a) and (b) for the Kolmogorov and IK scalings respectively.

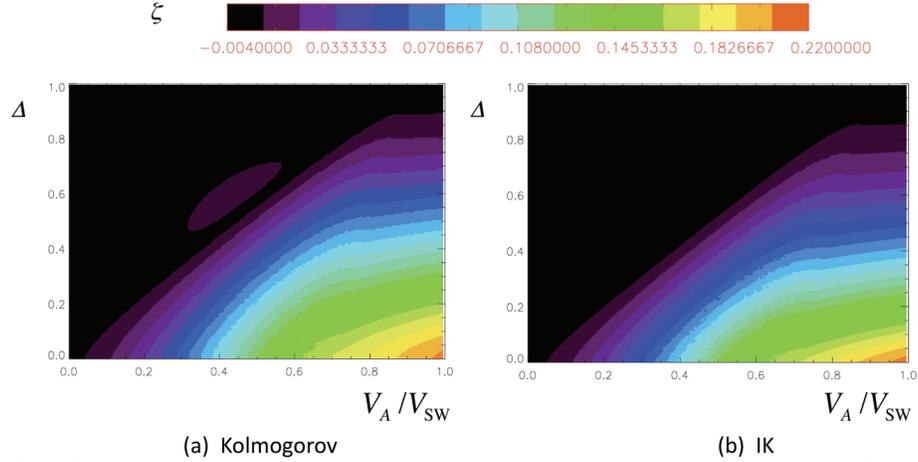

(a) Kolmogorov  (b) IK

**FIGURE 1.** Color-coded contour plots of the value of $\zeta$, calculated from Eqs. (1) and (2), for (a) the Kolmogorov scaling and (b) the IK scaling, as a function of $\Delta$ and $V_A/V_{SW}$.

From Fig. 1, it can be seen that $\zeta$ is larger for the Kolmogorov case. To see this point more quantitatively, we can compare its values along a line for a more realistic value of $V_A/V_{SW} = 0.075$ as a function of $\Delta$, as is shown in Fig. 2.

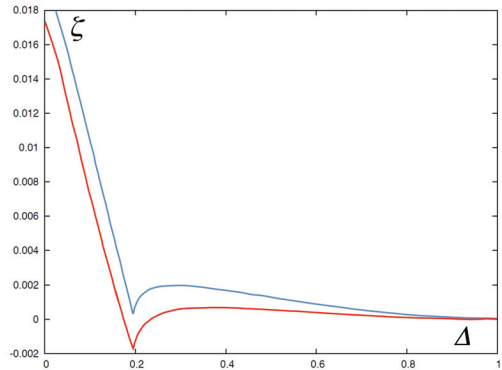

**FIGURE 2.** Value of $\zeta$, calculated from Eqs. (1) and (2), for the Kolmogorov scaling (blue) and the IK scaling (red), as a function of $\Delta$ for $V_A/V_{SW} = 0.075$.

This calculation shows that the effect of turbulence generation due to pickup ions is weaker when the IK spectrum is used. We can understand this by considering the physics of the $Q$ term, which indicates whether pickup ions give energy to a spectrum of waves (positive $Q$), or gain energy from it (negative $Q$). A pickup ion gives energy when it interacts with a backward moving wave at smaller wave number $k$, but gains energy when it interacts with a forward moving wave at larger $k$, due to the Doppler effect (see also [9,16]). The strength of such interactions is proportional to the intensity of the waves. For a turbulent spectrum of waves that decrease in intensity at larger values of *k*, pickup ions give energy to waves and result in a positive *Q* term. So, when the IK spectrum, which is flatter in *k* space, is used, there is a stronger cancellation between the two effects, resulting in smaller values of $\zeta$ and *Q*.

## COMPARISON WITH SOLAR WIND OBSERVATIONS

To see how a change of the pickup ions effect has on the solar wind heating prediction, we consider the case presented in [10], in which a more direct method is used in comparing predictions from the mode equations (1)-(3) with observations. Instead of setting the boundary conditions on $Z^2$, $\lambda$, and *T* at *r* = 1 AU in obtaining predictions for all *r*, the observed solar wind speed (which is assumed to be constant) at different positions *r* is used to determine when that fluid element actually passed through 1 AU. Then the solar wind conditions at that time at 1 AU are determined using Omnitape data, and used as boundary conditions for Eqs. (1)-(3). Here we follow their method, and repeat our study for the IK case, as shown in Fig. 3.

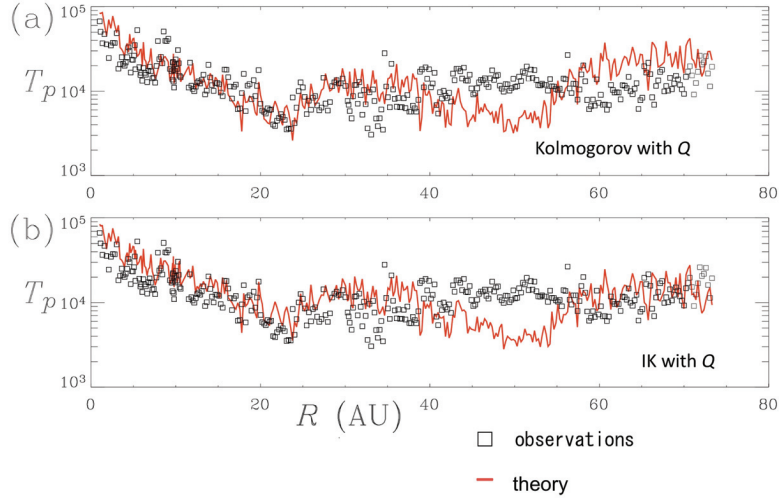

**FIGURE 3.** (a) The red curve is the solar wind proton temperature $T_p$ in K as a function of heliocentric distance in AU calculated from Eqs. (1)-(3) using the Kolmogorov scaling for the parameters used in [10]. The discrete data points are from *Voyager 2* observations. (b) The $T_p$ curve is now calculated using the IK scaling for the same parameters

In Fig. 3(a), the red curve is the proton temperature $T_p$ in K as a function of heliocentric distance in AU, calculated from Eqs. (1)-(3) using the Kolmogorov scaling for the parameters used in [10]. The discrete data points are from *Voyager 2* observations. We see that the predictions from the model are consistent with observations until about 43 AU. From there to about 55 AU, the predictions are substantially lower than observations. This is identified by [10] as a latitude effect, since *Voyager 2* was at high latitude. The predictions beyond 55 AU are also found to be somewhat higher than observations (about a factor of two on the average). In 3(b), the $T_p$ curve is now calculated using the IK scaling for the same parameters. We see that the agreement with observations up to about 43 AU is about the same as in the case (a). At the same time, the predictions beyond 55 AU are now lower, consistent with observations. However, the discrepancy with data from 43 to 55 AU is worse. Since the main discrepancy in this region is due to the high-latitude effect, it is hard to separate out the effects due to turbulence spectral laws.

## CONCLUSION

We have studied the effect of turbulence scaling laws on the heating of solar wind by substituting the IK cascade rate into a solar wind turbulence evolution model, replacing the Kolmogorov cascade rate, and comparing with observations from *Voyager 2* on the solar wind temperature. We show that the effect of pickup ions is weaker when using the IK spectrum instead. The solar wind temperature predicted by using the IK cascade is comparable with that using the Kolmogorov cascade, and seems to give even better agreement at the outer heliosphere.

## ACKNOWLEDGMENTS

This research is supported in part by National Science Foundation Grant No. AST-0434322, National Aeronautics and Space Administration Grant No. NNX08BA71G, and the Department of Energy under the auspices of the Center for Magnetic Self-Organization (CMSO) and the Center for Integrated Computation and Analysis of Reconnection and Turbulence (CICART).